\documentclass[10pt,journal]{IEEEtran}
\usepackage{setspace,amsmath,latexsym,cite,amssymb,epsfig,amsfonts,psfig}
\usepackage{url,cite}
\usepackage{balance}
\usepackage{graphicx}
\usepackage{enumerate}
\usepackage{stmaryrd}
\usepackage{colortbl}
\usepackage{url,cite}
\usepackage{balance}
\usepackage{graphicx}
\usepackage{enumerate}
\usepackage{multirow}
\usepackage{multicol}
\usepackage{slashbox}
\usepackage{booktabs}
\usepackage[table]{xcolor}
\definecolor{lightgray}{gray}{0.9}
\definecolor{lightgrayd}{gray}{1}

\usepackage{makeidx}  

\usepackage{amsthm}

        \makeatletter
        \def\fps@eqnfloat{!t}
        \def\ftype@eqnfloat{4}
        
        \newenvironment{eqnfloat*}
               {\@dblfloat{eqnfloat}}
               {\end@dblfloat}
        \makeatother
\ifCLASSOPTIONcompsoc
\else
\fi
\ifCLASSINFOpdf
\else
\fi

\begin{document}
\title{User Collusion Avoidance Scheme for Privacy-Preserving Decentralized Key-Policy Attribute-Based Encryption -- Full Version }
\author{ Yogachandran Rahulamathavan,~\textit{Member, IEEE}
\IEEEcompsocitemizethanks{
\IEEEcompsocthanksitem Y. Rahulamathavan is with the School of Engineering and Mathematical Science, City University London, London, U.K. (e-mail: Yogachandran.Rahulamathavan.1@city.ac.uk).\protect
\IEEEcompsocthanksitem Shorter version of this paper got accepted in IEEE TC \cite{Rahul9}. \protect
}
}
\IEEEcompsoctitleabstractindextext{
\begin{abstract}
Recent trend towards cloud computing paradigm,  smart devices and 4G wireless technologies has enabled seamless data sharing among users. Cloud computing environment is distributed and untrusted, hence data owners have to encrypt their data to enforce data confidentiality. The data confidentiality in a distributed environment can be achieved by using attribute-based encryption  technique. Decentralized attribute-based encryption technique is a variant of multiple authority based attribute-based encryption whereby any attribute authority  can independently join and leave the system without collaborating with the existing attribute authorities.  In this paper, we propose a privacy-preserving decentralized key-policy attribute-based encryption scheme. The scheme preserves the user privacy when users interact with multiple authorities to obtain decryption keys while mitigating the well-known user collusion security vulnerability. We showed that our scheme relies on decisional bilinear Diffie-Hellman standard complexity assumption in contrast to the previous nonstandard complexity assumptions such as $q-$decisional Diffie-Hellman inversion.
\end{abstract}
}
\maketitle

\IEEEdisplaynotcompsoctitleabstractindextext
\IEEEpeerreviewmaketitle


\section{Introduction}

Recent trend towards cloud computing, outsourcing, smart devices, and high bandwidth mobile broadband has enabled users to share information anywhere and anytime. Data could be shared using public data storages such as cloud computing infrastructure which can provide flexible computing capabilities at reduced costs. Although this brings many benefits, there are many unavoidable security problems since those facilities are maintained by third party service providers \cite{SRuj}.

Traditionally,  data owners, users, and  storage server are in the same  domain and  the storage server is fully trusted \cite{RBAC, CWDRBAC, ERBAC, TRBAC, SRBAC, ESRBAC, LRBAC, LAAC, ESTAC}. However, in cloud computing and outsourcing  environments, data confidentiality is not guaranteed since the data is stored and processed within the third party environment \cite{Rahul5,Rahul6,Rahul7,Rahul8}. Personnel information of the data owners and  commercial secrets of service providers can be leaked to third party if the data owners  store decrypted data in public servers. To overcome this challenge, the data confidentiality in a distributed environment can be achieved by applying attribute-based encryption (ABE) technique \cite{FIBE, kpabe,cpabe}. ABE  is considered as a promising cryptographic technique for organizations who host their services in the cloud \cite{FIBE, kpabe,cpabe,Rahul1}. In fact, using ABE, the data owners can enforce fine-grained access policies based on the nature of data. In ABE, at least four different parties are involved namely data owner, users, attribute authority and storage server (i.e., cloud). 

The attribute authority is responsible for a set of attributes and will be issuing encryption and decryption credentials to the data owner and users, respectively. In order to get the decryption credentials for a set of attributes, the user needs to prove to the attribute authority that he is a legitimate user for those attributes. Data owner will choose a set of attributes for a particular data and will use the corresponding encryption credential for those attributes to encrypt the data. Then the encrypted data will be uploaded onto the cloud by the data owner. If the user wants to access the encrypted data, then he will first download the encrypted data from the cloud and will compare whether he satisfies the set of attributes defined by the data owner during the encryption. For instance, let us assume, an employer uploads an  encrypted file to the cloud using ABE, where the access policy of that file is defined using the following attributes and functions AND and OR: ``Manager" OR ``Finance Office" AND ``Company A". Hence, an employee who is a ``Manager"  employed at ``Company A" can  decrypt the file.

\begin{table*}[!ht]\caption{Comparisons of attribute based encryption schemes related to the proposed scheme.}
\begin{tabular}{|c|c|c|c|c|c|c|}
\hline
Ref.                        & Year & KP/CP-ABE & \begin{tabular}[c]{@{}c@{}}Access\\ Structure\end{tabular} & Security Model     & \begin{tabular}[c]{@{}c@{}}Decentralized\\ Attribute Authorities\end{tabular} & \begin{tabular}[c]{@{}c@{}}User\\ Privacy\end{tabular} \\ \hline
Sahai et al. {[}17{]}       & 2005 & KP-ABE    & Threshold                                                  & Selective Security & No                                                                            & No                                                     \\ \hline
Goyal et al. {[}18{]}       & 2006 & KP-ABE    & Tree                                                       & Selective Security & No                                                                            & No                                                     \\ \hline
Chase {[}20{]}              & 2007 & KP-ABE    & Threshold                                                  & Selective Security & No                                                                            & No                                                     \\ \hline
Bethencourt et. al {[}19{]} & 2007 & CP-ABE    & Tree                                                       & Selective Security & No                                                                            & No                                                     \\ \hline
Chase et al. {[}21{]}       & 2009 & KP-ABE    & Threshold                                                  & Selective Security & No                                                                            & Yes                                                    \\ \hline
Lewko et al. {[}23{]}       & 2011 & CP-ABE    & LSSS                                                       & Selective Security & Yes                                                                           & No                                                     \\ \hline
Liu et al. {[}28{]}         & 2011 & CP-ABE    & LSSS                                                       & Full Security      & Yes                                                                           & No                                                     \\ \hline
Han et al. {[}15{]}         & 2012 & KP-ABE    & Tree                                                       & Selective Security & Yes                                                                           & Yes                                                    \\ \hline
Our Scheme                  & 2015 & KP-ABE    & Tree                                                       & Selective Security & Yes                                                                           & Yes                                                    \\ \hline
\end{tabular}
\end{table*}

There are two main types of ABE namely ciphertext-policy ABE (CP-ABE) and key-policy ABE (KP-ABE) \cite{kpabe,cpabe}. In CP-ABE scheme, the access structure is assigned to the ciphertext and each private key is associated with a set of attributes. It must be noted that in CP-ABE, the data owner must enforce the access structure together with encryption. The user must have the private key to satisfy the policy in order to decrypt the ciphertext \cite{cpabe}. In the KP-ABE scheme, the private keys are associated with an access structure and the ciphertext is labeled with a set of attributes. When the access structure defined in the private key matches the attributes labeled with the ciphertext, then it decrypts the ciphertext \cite{kpabe}. Both the schemes have different applications in secure data sharing. However, we consider only KP-ABE scheme in this paper.

Each ABE scheme can be further divided into  two types: single authority based ABE \cite{kpabe} and multiple authorities based ABE (MA-ABE)  \cite{imaabe} schemes. In a single authority based ABE scheme, only one attribute authority is responsible for monitoring all the attributes. In the MA-ABE scheme, in contrast to the single authority ABE scheme, there are multiple attribute authorities responsible for a disjoint sets of attributes. The management of attributes is a crucial part in the ABE systems.  In reality, it is more convenient and secure to monitor and maintain different sets of attributes  by different attribute authorities, e.g., in healthcare one authority can monitor attributes of nurse and doctors while another authority monitors attributes of  administrators and human resources \cite{Rahul2} or in vehicular adhoc network (VANET), different identities can be monitored by different authorities \cite{Rahul3, Rahul4}. The focus of this paper is on multi-authority KP-ABE scheme. Let us review some of the relevant works in multi-authority ABE schemes followed by the decentralized schemes.

\textbf{Related Works:} Chase et. al. \cite{maabe} presented a MA-ABE scheme, which allows any polynomial number of independent authorities to monitor attributes and distribute decryption keys. In \cite{maabe}, the data owner chooses a number, i.e., $m_{k}$ for $k${th} attribute authority, and a set of attributes from each attribute authority, and encrypts a message. This encrypted message can be decrypted  only by the users who satisfy  $m_{k}$ number of attributes from the $k{th}$ attribute authority, $k=1,\ldots,N$. However,  a trusted central authority  is needed for distributing all the keys  \cite{maabe}.

An improved MA-ABE scheme without central authority has been presented by Chase and Chow in \cite{imaabe} where, each pair of attribute authorities securely exchange a shared secret among them during the set up process. In \cite{maabe}, users must submit their global identity (GIDs) to each authority to obtain the decryption credentials. This will breach the user privacy since a set of corrupted authorities can pool together all the attributes belong to the particular GID. In order to mitigate this privacy vulnerability, Chase and Chow proposed an anonymous key issuing protocol in \cite{imaabe}, whereby a user can obtain the decryption keys from attribute authorities without revealing her GIDs. Even though the scheme proposed by Chase and Chow eliminates the central authority, all the attribute authorities must be online and collaborate with each other to set up the ABE system. Hence, if one attribute authority joins and/or leaves the system then the entire system must be rebooted. Let us review the decentralized MA-ABE schemes in literature.

In literature, various protocols  have been proposed to decentralize  the ABE scheme, however, each scheme has its own merits and demerits. In a decentralized ABE scheme, any attribute authority  can independently join and/or leave the system or  issue keys to users without collaborating with existing attribute authorities. Hence, this kind of scheme eliminates the need that all the attribute authorities must be online and interact with each other in order to setup the system.

Decentralized schemes for CP-ABE was proposed in \cite{Dabe,Muller1, Muller2, ZLiu, OT13}.  In particular, \cite{Dabe} constructs the first decentralized multi-authority ABE scheme whereby any party can become an authority and there is no requirement for any global coordination
other than the creation of an initial set of common reference parameters. However, the schemes in \cite{Dabe,BWaters,Muller1, Muller2, ZLiu, OT13} do not preserve the user privacy i.e., attributes of users can be collected by tracing users' GIDs. For the first time, Han et al. proposed a privacy-preserving decentralized scheme for KP-ABE \cite{main1}. In contrast to the existing decentralized ABE schemes,  the scheme in \cite{main1} preserves the user privacy and relies only on decisional bilinear Diffie-Hellman (DBDH) standard complexity assumption.
In \cite{main1}, the GID of the user is used to tie all the decryption keys together, where blind key generation protocol has been used to issue the decryption keys. Hence, corrupted attribute authorities cannot pool the users' attributes by tracing the GIDs' of the users from the decryption keys. Unfortunately, the scheme in \cite{main1} cannot prevent  user collusion, hence, two users can pool their decryption keys  to generate decryption keys for an unauthorized user \cite{main2}. This is due to weak bind between users' GID and the decryption keys.

In this paper, contrasting to all the works in literature, we propose a strong privacy-preserving decentralized KP-ABE scheme in order to mitigate the known user collusion security vulnerability \cite{main2}. We exploit  the anonymous key issuing protocol in \cite{imaabe} to strengthen the bind between decryption keys and GID as well as to preserve the user privacy. In order to incorporate the anonymous key issuing protocol, we  modify the privacy-preserving decentralized KP-ABE scheme in \cite{main1}. We prove this by contradiction that the proposed scheme is secure i.e., we reduce the DBDH standard complexity assumption to show that an adversary who can break the proposed scheme  can be exploited to break the DBDH assumption. We also proved that the anonymous key issuing protocol is free from leak and selective-failure.

The reminder  of this paper is organized as follows: we provide notations and cryptographic building blocks required for the new algorithm in Section II. We propose the decentralized KP-ABE algorithm and it's security proof in Section III. In Section IV, we describe anonymous key issuing protocol and it's security proof. Complexities of similar algorithms are compared in Section V followed by Conclusions  in Section VI.

\section{Preliminaries}
The following notations will be used throughout this paper. We use $x~\underleftarrow{~~R~~}~X$ to denote that $x$ is randomly selected from $X$. Suppose $\mathbb{Z}_p$ is a finite field with prime order $p$, by $\mathbb{Z}_p[x]$, we denote the polynomial ring on $\mathbb{Z}_p$. Let us explain the building blocks used in this paper in the following subsections.

\subsection{Lagrange Interpolation}
Shamir's secret share uses Lagrange interpolation technique to obtain the secret from shared-secrets. Suppose that $p(x) \in \mathbb{Z}_p[x]$ is a $(k-1)$ degree polynomial and secret $s = p(0)$. Let us denote $S=\{x_1, x_2, \ldots, x_k\}$ and the Lagrange coefficient for $x_i$ in $S$ as
\begin{equation*}
    \Delta_{x_i, S}(x)=\prod\limits_{x_j~\in~S,~x_j~\neq x_i}\frac{x-x_j}{x_i-x_j}.
\end{equation*}
For a given $k$ different number of values $p(x_1)$, $p(x_2)$, $\ldots$, $p(x_k)$, the polynomial $p(x)$ can be reconstructed as follows,
\begin{eqnarray*}
    p(x)&=&\sum_{x_i~\in~S}p(x_i)\prod\limits_{x_j~\in~S,~x_j~\neq x_i}\frac{x-x_j}{x_i-x_j},\\
    &=&\sum_{x_i~\in~S}p(x_i)\Delta_{x_i, S}(x),
\end{eqnarray*}
hence the secret $s$ can be obtained as:
\begin{eqnarray*}
    \nonumber s&=&p(0),\\
    &=&\sum_{x_i~\in~S}p(x_i)\prod\limits_{x_j~\in~S,~x_j~\neq x_i}\frac{0-x_j}{x_i-x_j}.
\end{eqnarray*}

\subsection{Bilinear Groups}
Let $\mathbb{G}_{1}$, $\mathbb{G}_{2}$ be two multiplicative groups of prime order $p$, generated by  ${g}_{1}$, ${g}_{2}$ respectively.
A bilinear map is denoted as $\hat{e}$ : $\mathbb{G}_{1}$ $\times$ $\mathbb{G}_{1}$ $\rightarrow \mathbb{G}_{{2}}$, where it has the following three properties.
\begin{enumerate}
  \item Bilinearity: $\forall {x}, y \in \mathbb{G}_{{1}}$, and ${a}, {b} \in \mathbb{Z}_{\textit{p}}$, there is $\hat{e}({x}^{a}, {y}^{b}) = \hat{e}(x, y)^{ab}$.
  \item Non-degeneracy:  $\hat{e}({g}_{1}, {g}_{1}) \neq 1$ where $1$ is the identity of $\mathbb{G}_{{2}}$.
  \item Computability: $\hat{e}$ is an efficient computation.
\end{enumerate}

\subsection{Decisional Bilinear Diffie-Hellman Complexity Assumption}
The DBDH  is one of the standard cryptographic complexity assumption in contrast to other nonstandard complexity assumptions such as $q-$decisional Diffie-Hellman Inversion. Let $a, b, c, z ~\underleftarrow{~~R~~} \mathbb{Z}_p$, $(\hat{e}, p, \mathbb{G}_{1}, \mathbb{G}_{2})$ as bilinear parameters, and $g$ as a generator of $\mathbb{G}_{1}$. The DBDH assumption says that in $(\hat{e}, p, \mathbb{G}_{1}, \mathbb{G}_{2})$, there is no probabilistic polynomial-time adversary can distinguish [$g^a, g^b, g^c, \hat{e}(g,g)^{abc}$]  from [$g^a, g^b, g^c, \hat{e}(g,g)^{z}$] with non-negligible advantage. We will use this property to prove by contradiction  that our proposed algorithm is secure against well-known attacks. Later in this paper, we will show that if there is an adversary who can break the proposed algorithm then we can use the adversary indirectly to break the DBDH assumption (i.e., this is a contradiction to the DBDH assumption, hence our proposed algorithm is secure).

\subsection{Commitment}
A commitment scheme consists of three sub-algorithms: \textit{Setup}, \textit{Commit}, and \textit{Decommit}. The \textit{Setup} outputs parameters (i.e., \textit{params}) for a particular security input. The \textit{Commit} outputs commitment (e.g., \textit{com}) and decommitment (e.g., \textit{decom}) parameters for a particular message  and \textit{params}. The \textit{Decommit} algorithm outputs $1$ if the \textit{com} matches the \textit{decom} for the message and \textit{params}, otherwise it outputs $0$.

\subsection{Proof of Knowledge}
We use the notation introduced in \cite{PoK} to prove statements about discrete logarithm. By
\begin{equation*}
    PoK\left\{(\alpha,\beta,\gamma):y=g^\alpha h^\beta \bigwedge \tilde{y}=\tilde{g}^\alpha \tilde{h}^\beta\right\}
\end{equation*}
we denote a zero knowledge proof of knowledge of integers $\alpha,\beta$ and $\gamma$. Conventionally, the values in the parenthesis
denote the knowledge that is being proven, while the rest of the other values are known to the verifier. There exists a knowledge extractor which can be used to rewind these quantities from a successful prover.

\subsection{Access Structure}
Let $\{att_1, att_2, \ldots, att_n\}$ be a set of attributes. A collection $\mathbb{A} \subseteq 2^{\{att_1, att_2,\ldots,att_n\}}$ is monotone if $\forall B, C :$ if $B \in A$ and $B \subseteq C$ then $C \in A$. An access structure (respectively, monotone access structure) is a collection (respectively, monotone collection) $\mathbb{A}$ of non-empty subsets of $\{att_1, att_2, \ldots, att_n\}$, i.e., $\mathbb{A} \subseteq 2^{\{att_1, att_2,\ldots,att_n\}}/ \{\phi\}$. The sets in $\mathbb{A}$ are called the authorized sets, and the sets not in $\mathbb{A}$ are called the unauthorized sets \cite{Beimel}. In our context, we restrict our attention to monotone access structures. However, it is also possible to (inefficiently) realize general access structures using the techniques given in \cite{kpabe}.

We follow the same strategy used in \cite{kpabe} to build monotone access structure using access tree. Let $\mathbb{T}$ be a tree representing an access structure. Each non-leaf node of the tree represents a threshold gate, described by its children and a threshold value. If $num_x$
is the number of children of a node $x$ and $k_x$ is its threshold value, then $0 < k_x = num_x$. When $k_x = 1$, the threshold gate is an OR gate and when $k_x = num_x$, it is an AND gate.

Each leaf node $x$ of the tree is described by an attribute and a threshold value $k_x = 1$.
To facilitate working with the access trees, we define a few functions. We denote the
parent of the node $x$ in the tree by $parent(x)$. The function $att(x)$ is defined only if $x$ is a
leaf node and denotes the attribute associated with the leaf node $x$ in the tree. The access
tree $\mathbb{T}$ also defines an ordering between the children of every node, that is, the children
of a node are numbered from $1$ to $num$. The function $index(x)$ returns such a number
associated with the node $x$. Where the index values are uniquely assigned to nodes in the
access structure for a given key in an arbitrary manner.

Let $\mathbb{T}$ be an access tree with root $r$. Denote by $\mathbb{T}_x$ the subtree
of $\mathbb{T}$ rooted at the node $x$. Hence $\mathbb{T}$ is the same as $\mathbb{T}_r$. If a set of attributes $\gamma$ satisfies the
access tree $\mathbb{T}_x$, we denote it as $\mathbb{T}_x(\gamma) = 1$. We compute $\mathbb{T}_x(\gamma)$ recursively as follows. If $x$ is a
non-leaf node, evaluate $\mathbb{T}_x'(\gamma)$ for all children $x'$
of node $x$. $\mathbb{T}_x(\gamma)$ returns 1 if and only if at
least $k_x$ children return 1. If $x$ is a leaf node, then $\mathbb{T}_x(\gamma)$ returns $1$ if and only if $att(x) \in \gamma$.


\section{Decentralized Key-Policy Attribute-based Encryption}
In this section, we present our decentralized KP-ABE scheme.   In a decentralized scheme, it is not necessary to maintain a fixed number of attribute authorities. Any attribute authority can join and/or leave the system at any time without rebooting the system. First of all, we will explain  sub-algorithms and security game followed by the privacy-preserving decentralized KP-ABE. In Section IV, we incorporate the anonymous key issuing protocol to strengthen the bind between user GID and decryption keys, hence, our scheme mitigates the user collusion vulnerability found in  \cite{main1}.
\begin{figure*}\small
\setlength{\unitlength}{0.2in} 
\centering 
\framebox{
\parbox[t][12cm]{17cm}{
\textbf{Global Setup} $\mathcal{GS}$
\begin{itemize}
  \item For a given security parameter $\lambda$, $\mathcal{GS}$ generates the bilinear groups $\mathbb{G}_1$ and $\mathbb{G}_2$ with prime order $p$ as follows:
      ${\{\mathbb{G}_{1}, \mathbb{G}_{2}\} \leftarrow \mathcal{GS}(1^{\lambda })}$
  \item Let $e: \mathbb{G}_{1}\times\mathbb{G}_{1}\longrightarrow\mathbb{G}_{2}$ be a bilinear map and $g$, $h$ and $h_1$ be the generators of $\mathbb{G}_{1}$ such that $\forall~x, y~\in~\mathbb{G}_{1}$ and $\forall~a,~b~\in~\mathbb{Z}_p$, $e(x^a,y^b)=e(x,y)^{ab}$
  \item There are $N$ number of authorities $\{A_1,\ldots,A_N\}$: $A_k$ monitors $n_k$ attributes i.e. $\tilde{A}_k=\{a_{k,1}\ldots,a_{k,n_k}\}, \forall k$
\end{itemize}

\textbf{Authorities Setup} $\mathcal{AS}$
\begin{itemize}
  \item  Security parameters of $A_k$:
   $SK_k = \left\{ \alpha_{k}, \beta_k,~ \textrm{and}~ [t_{k,1},\ldots,t_{k,n_k}]\right\} ~\underleftarrow{~~R~~} \mathbb{Z}_{p}, \forall k$
  \item Public parameters of $A_k$: $PK_k$ = \{$Y_{k} = e(g, g)^{\alpha_k}$, $Z_{k} = g^{\beta_k}$, and $[T_{k,1}=g^{t_{k,1}},\ldots,T_{k,n_k}=g^{t_{k,n_k}}]\}$, $\forall$ $k$
  \item $A_k$ specify $m_k$ as  minimum number of attributes required to satisfy the access structure ( $m_k \leq n_k$)
\end{itemize}

\textbf{Key Generation} $\mathcal{KG}$
\begin{itemize}
    \item Collision-Resistant Hash Function $H: \left \{ 0,1 \right \}^{*} \rightarrow \mathbb{Z}_{p}$ to generate  $u$ from the user global identity
    \item Attribute set of user is $\tilde{A}_u$: $\tilde{A}_u\cap \tilde{A}_k=\tilde{A}_u^k ~\forall ~k$
    \item $A_k$ generates $r_{k,u} \in _{R}  \mathbb{Z}_{p}$ and  polynomial $q_x$ for each node $x$ (including the leaves) $\mathbb{T}$
    \item For each node $x$, the degree $d_x$ of the polynomial $q_x$ is $d_x = k_x - 1$  where $k_x$ --  threshold value of that node
    \item Now, for the root node $r$, set $q_r(0) = r_{k,u}$
    \item For any other node $x$, set $q_x(0) = q_{parent(x)}(index(x))$
    \item Now decryption keys for the user $u$ is generated as follows:\\
    \ \\
    $~~~~~~~~D = \left[D_{k,u}=g^{-\alpha_k}h^{\frac{\beta_k}{r_{k,u}+u}}h_1^{\frac{r_{k,u}}{\beta_k+u}}, D_{k,u}^1=h^{\frac{1}{r_{k,u}+u}}, D_{k,u}^j=h_1^{\frac{q_{a_{k,j}}(0)}{(\beta_k+u)t_{k,j}}}, \forall a_{k,j} \in \tilde{A}_u^k\right]$
\end{itemize}

\textbf{Encryption} $\mathcal{E}$
\begin{itemize}
\item Attribute set for the message $m$ is $\tilde{A}_m$: $\tilde{A}_m\cap \tilde{A}_k=\tilde{A}_m^k, \forall k$, i.e. $\tilde{A}_m= \{ \tilde{A}_m^1,\ldots,\tilde{A}_m^k,\ldots,\tilde{A}_m^N \}$
\item Data owner of message $m$ randomly chooses $s \in _{R} \mathbb{Z}_{p}$, and output the ciphertext as follows:

$~~~~C = \left [ C_1 = m.\prod\limits_{k \in I_C}e(g, g)^{\alpha_ks}, C_2= g^{s}, C_3=\prod\limits_{k \in I_C}g^{\beta_ks} \textrm{and}
\left \{ C_{k,j} = T_{k,j}^{s} \right \}_{\forall k\in I_{C}, a_{k,j} \in \tilde{A}_m^j} \right]$ where $I_c$ denotes the index set of the authorities.
\end{itemize}

\textbf{Decryption} $\mathcal{D}$
\begin{itemize}
  \item In order to decrypt $C$, the user $u$, computes $X,~Y$ and $Q_k$ as follows:
\item[--]   $\left[X = \prod\limits_{k \in I_C} e\left(C_2,D_{k,u}\right),~Y =  e\left(C_3,D_{k,u}^1\right) \textrm{and}~S_k=\prod\limits_{a_{k,j} \in A_m^k}e\left(C_{k,j}, D_{k,u}^j\right)^{\Delta_{a_{k,j} ,\tilde{A}_m^j}(0)}\right]$
  \item User then decrypts the message $m$ as follows: $m=\frac{C_1X}{Y\prod\limits_{k \in I_C}S_k}$
\end{itemize}
}}
\caption{The proposed decentralized key-policy attribute-based encryption scheme.}
\label{main algorithm}
\end{figure*} 

\subsection{Sub-algorithms}
Our algorithm contains five sub-algorithms namely global setup, authority setup, key issuing, encryption and decryption. Let us briefly explain the functionalities of each sub-algorithm.

\noindent\textbf{Global Setup:} This algorithm takes a security parameter as input and output system parameters. These system parameters can be used by authorities who join the system.

\noindent\textbf{Authority Setup:} Each attribute authority uses the system parameters obtained from the global setup to generate public and private keys for the attributes it maintains.

\noindent\textbf{Key Issuing:}  User and  attribute authority interact via anonymous key issuing protocol (see Section IV) in order to determine a set of attributes belongs to the user. Then attribute authority  generates decryption credentials for  those attributes and send them to the user.

\noindent\textbf{Encryption:} The encryption algorithm takes a set of attributes maintained by attribute authority and the data as input. Then it outputs the ciphertext of the data.

\noindent\textbf{Decryption:} The decryption algorithm takes  the decryption credentials received from attribute authorities and the ciphertext as input. The decryption will be successful if and only if the user attributes satisfy the access structure.

\subsection{Security Game}
In order to avoid the security vulnerabilities, ABE schemes should be proven to be secure against the selective identity (ID) model \cite{FIBE}. In the selective ID model, the adversary should provide the identities of the attribute authorities (challenge identities) he wishes to challenge the challenger with. Then challenger (i.e., the system) will generate necessary parameters corresponding to the challenge identities  and send them to the adversary. Then the adversary is allowed to make secret queries about the challenge identities. If the adversary cannot decrypt the encrypted message at the end with non-negligible advantage then the proposed scheme is secure against the selective ID model. Formally, this is represented by the following game between the adversary and the challenger:\\

\noindent\textbf{Setup}
\begin{itemize}
  \item Adversary sends a list of attribute sets and attribute authorities including corrupted authorities to the challenger.
  \item Now the challenger generates public and private keys corresponding to the attributes and authorities provided by the adversary.
  \item Challenger provides public and private keys corresponding to the corrupted authorities to the adversary while only public keys corresponding to the remaining authorities to the adversary.
\end{itemize}
\noindent\textbf{Secret Key Queries}
\begin{itemize}
  \item The adversary is allowed to make any number of secret key queries as he wants against the attribute authorities.
  \item However, the only requirement is that for each user, there must be at least one non corrupted attribute  authority from which the adversary can get insufficient number of secret keys.
\end{itemize}
\noindent\textbf{Challenge}
\begin{itemize}
  \item The adversary sends two messages $m_0$ and $m_1$ to the challenger in plain domain.
  \item Now the challenger randomly chooses one of the messages and encrypt it and send the ciphertext to the adversary.
\end{itemize}
\noindent\textbf{More Secret Key Queries}
\begin{itemize}
  \item The adversary is allowed to make more secret key queries as long as he satisfies the requirement  given earlier.
\end{itemize}
\noindent\textbf{Guess}
\begin{itemize}
  \item Now the adversary guesses which message was encrypted by the challenger.
  \item The adversary is said to be successful if he guesses the correct message with probability $\frac{1}{2}+\epsilon$ whereby $\epsilon$ is a non-negligible function.
\end{itemize}

\subsection{Construction of our new algorithm}
Let us consider a system which contains $N$ number of attribute authorities (i.e., we denote them as $A_1,\ldots,A_N$). The attribute set managed by the authority $A_k$ is denoted as  $\widetilde{A_k}=\{a_{k,1}, \ldots, a_{k,n_k}\}$ $\forall k$. Each attribute authority also assigned a value $d_k$ i.e., user must have at least $d_k$ number of attributes of this authority in order to retrieve the secret key associated with this attribute authority. The complete algorithm is given in Fig.~\ref{main algorithm}. Let us explain the important steps involved in Fig.~\ref{main algorithm}.

Initially, for a given security parameter $\lambda$, global setup algorithm ($\mathcal{GS}$) generates the bilinear groups $\mathbb{G}_1$ and $\mathbb{G}_2$ with prime order $p$ i.e., ${\{\mathbb{G}_{1}, \mathbb{G}_{2}\} \leftarrow \mathcal{GS}(1^{\lambda })}$. The authority setup algorithm ($\mathcal{AS}$) is executed by each attribute authority to randomly generate public keys ($PK$) and the corresponding secret keys ($SK$).  The public-secret key pairs for $A_k$ is given as $\left\{(Y_{k}, Z_k, [T_{k,1},\ldots,T_{k,n_k}]), (\alpha_{k}, \beta_k, [t_{k,1},\ldots,t_{k,n_k}])\right\}$.

Let us denote the attribute set belongs to user $u$ as $\widetilde{A_u}$ and the common attribute set between user $u$ and authority $k$  as $\widetilde{A_u^k}$ i.e., $\widetilde{A_u^k} = \widetilde{A_u} \bigcap \widetilde{A_k}$. Key generation ($\mathcal{KG}$) algorithm will be used to issue decryption keys to the user $u$ with a set of attributes $\widetilde{A_u}$. The algorithm outputs a key that enables the user to decrypt a message encrypted under a set of attributes $\widetilde{A_u^k}$ if and only if $\mathbb{T} (\widetilde{A_u^k})= 1$.  The algorithm proceeds as follows. First choose a polynomial $q_x$ for each node $x$ (including the leaves)
in the tree $\mathbb{T}$. These polynomials are chosen in the following way in a top-down manner,
starting from the root node $r$. For each node $x$ in the tree, set the degree $d_x$ of the polynomial $q_x$ to be one less than
the threshold value $k_x$ of that node, that is, $d_x = k_x - 1$. Now, for the root node $r$, set
$q_r(0) = r_{k,u}$ and $d_r$ other points of the polynomial $q_r$ randomly to define it completely. For
any other node $x$, set $q_x(0) = q_{parent(x)}(index(x))$ and choose $d_x$ other points randomly to
completely define $q_x$. Once the polynomials have been decided, for each leaf node $x$, we give the
secret value  $D_{k,u}$, $D_{k,u}^1$ and $D_{k,u}^j$, $\forall a_{k,j} \in \widetilde{A_u^k}$ as shown in Fig.~\ref{main algorithm} to the user.
Note that, users' unique identifier $u$ is tied non-linearly with each decryption key to strengthen the bind, hence the vulnerability found in \cite{main1} can be mitigated with the proposed approach. 

Let us denote the set of attributes used to encrypt  message $m$ as $\widetilde{A_m}$ and the common attribute set between message $m$ and the authority $k$ as $\widetilde{A_m^k}$ i.e., $\widetilde{A_m} = \{\widetilde{A_m^1},\ldots,\widetilde{A_m^k},\ldots,\widetilde{A_m^N}\}$. Let us also denote the index set of authorities involved in the ciphertext of message $m$ as $I_c$. The encryption algorithm ($\mathcal{E}$)  encrypts the message $m \in \mathbb{G}_2$ using an attribute set $\widetilde{A_m}$. In order to encrypt the message, the message owner randomly generates $s$ and computes ciphertext $C=\left[C_1, C_2, C_3, C_{k,j}, \forall a_{k,j} \in \widetilde{A_m^k}\right]$. If user has decryption keys for the attributes of message $m$ then he can obtain the message $m$ from the ciphertext using the following four steps by executing the decryption algorithm ($\mathcal{D}$). First, user can use decryption key $D_{k,u}$ and $C_2$ to compute $X$ as
\begin{eqnarray*}
    X\!\!\!\!\!&=&\!\!\!\!\!\!\!\prod\limits_{k \in I_C} e\left(C_2,D_{k,u}\right),\\
    &=&\!\!\!\!\!\!\!\prod\limits_{k \in I_C} e\left(g^s,g^{-\alpha_k}h^{\frac{\beta_k}{r_{k,u}+u}}h_1^{\frac{r_{k,u}}{\beta_k+u}}\right),\\
    \!\!\!\!\!&=&\!\!\!\!\!\!\!\prod\limits_{k \in I_C}e\left(g,g\right)^{-s\alpha_k}\prod\limits_{k \in I_C}e\left(g,h\right)^{\frac{s\beta_k}{r_{k,u}+u}}\prod\limits_{k \in I_C}e\left(g,h_1\right)^{\frac{sr_{k,u}}{\beta_k+u}},
\end{eqnarray*}
then user uses decryption key $D_{k,u}^1$ and $C_3$ to compute $Y$ as
\begin{eqnarray*}
    Y &=&  e\left(C_3,D_{k,u}^1\right),\\
      &=&  e\left(\prod\limits_{k \in I_C}g^{\beta_ks},h^{\frac{1}{r_{k,u}+u}}\right),\\
    & = &\prod\limits_{k \in I_C}e\left(g,h\right)^{\frac{s\beta_k}{r_{k,u}+u}},
\end{eqnarray*}
and uses $D_{k,u}^j$, $C_{k,j}$, $\forall a_{k,j}\in \widetilde{A_m^j}$ and polynomial interpolation to get $r_{k,u}$ as

\begin{eqnarray*}
S_k&=&\prod\limits_{a_{k,j} \in \widetilde{A_m^k}}e\left(C_{k,j}, D_{k,u}^j\right)^{\Delta_{a_{k,j} \in \widetilde{A_m},\widetilde{A_m^j}}(0)},\\
   &=&\prod\limits_{a_{k,j} \in \widetilde{A_m^k}}e\left(g^{t_{k,j}s}, h_1^{\frac{q_{a_{k,j}}(0)}{(\beta_k+u)t_{k,j}}}\right)^{\Delta_{a_{k,j} \in \widetilde{A_m},\widetilde{A_m^j}}(0)},\\
   &=&\prod\limits_{a_{k,j} \in \widetilde{A_m^k}} e(g, h_1)^{\frac{s}{(\beta_k+u)}q_{a_{k,j}}(0)\Delta_{a_{k,j} \in \widetilde{A_m},\widetilde{A_m^j}}(0)},
\end{eqnarray*}
where $q_{a_{k,j}}(0)=q_{parent(a_{k,j})}(index(a_{k,j}))$. Hence,
\begin{eqnarray*}
 S_k&=& e(g, h_1)^{\frac{sr_{k,u}}{(\beta_k+u)}}.
\end{eqnarray*}
Now user can get the message $m$ using $C_1$ and pre-computed values $X, ~Y,~ S_k,~ \forall k$ as follows:

\begin{eqnarray*}
&&\!\!\!\!\!\!\!\!\!\!\!\! \frac{C_1X}{Y\prod\limits_{k \in I_C}S_k}=\left( \frac{m.\prod\limits_{k \in I_C}e(g, g)^{\alpha_ks}}{\prod\limits_{k \in I_C}e(g, h_1)^{\frac{s\beta_{k,u}}{r_k+u}}}\right)~ \times ~\\
&&\!\!\!\!\!\!\!\!\!\!\!\! \left( \frac{\prod\limits_{k \in I_C}e\left(g,g\right)^{-s\alpha_k}\!\!\!\! \prod\limits_{k \in I_C}e\left(g,h\right)^{\frac{s\beta_k}{r_{k,u}+u}} \!\!\!\! \prod\limits_{k \in I_C}e\left(g,h_1\right)^{\frac{sr_{k,u}}{\beta_k+u}}}{\prod\limits_{k \in I_C}e(g, h_1)^{\frac{sr_{k,u}}{\beta_k+u}}}\right),\\
&=& m.
\end{eqnarray*}


\subsection{Security Analysis}

\noindent \textbf{Theorem 1.} \emph{The proposed scheme is  semantically secure against chosen plain text attack (CPA) in the selective ID model, if there exist negligible function $\upsilon$ such that,  any adversary will succeed the security game explained earlier with probability at most $\frac{1}{2}+\upsilon$.}\\

\noindent \textbf{Proof.} Suppose if there is a probabilistic polynomial time adversary who can break our algorithm then  there will be a challenger who  can break the DBDH assumption by exploiting the adversary. Lets assume that the challenger is provided with  [$g^a, g^b, g^c, Z$] and if  the challenger wants to break the DBDH assumption then he needs to determine whether $Z=\hat{e}(g,g)^{abc}$  or not with at least $\frac{1}{2}+\upsilon$ probability.

Let us assume that there is an adversary who can break the proposed algorithm. In this section, we will show that the challenger can use such an adversary to break the DBDH assumption. In order to exploit such an adversary, the challenger needs to incorporate the given [$g^a, g^b, g^c, Z$] within the proposed algorithm (i.e., Fig.~\ref{main algorithm}). First of all, let us explain how the challenger incorporates [$g^a, g^b, g^c, Z$] within the global setup, authority setup, and key generation sub-algorithms. We stress here that this incorporation is indistinguishable from the steps provided in Fig.~\ref{main algorithm}.

Initially, as explained in the security game, the adversary must submit a set of attributes and a set of attribute authorities he wants to challenge. Let us denote the set of attributes provided by the adversary as $\widetilde{A_C}=\{\widetilde{A_C^1},\ldots, \widetilde{A_C^k}, \ldots,\widetilde{A_C^N}\}$ where $\widetilde{A_C^k}=\widetilde{A_C}\bigcap\widetilde{A_k}$ and a list of corrupted authorities as $C_A$. One of the conditions as given in security game is that at least there will be one honest authority for each user whereby the adversary can get insufficient number of decryption credentials \cite{maabe}. Let us denote GID of a particular user as $\omega$ and the corresponding honest authority as $\kappa$, and it's access structure $\mathbb{T}^{\kappa}$ hence $\mathbb{T}^{\kappa}(\widetilde{A_C^{\kappa}}\cap \widetilde{A_\omega^{\kappa}})=0$. Hence, we can divide the authorities who maintain attributes $A_C$ into three: corrupted authorities, authorities who are not corrupted and not $\kappa$, and $\kappa$.
Firstly, challenger generates two random values $\gamma, \eta ~\underleftarrow{~~R~~} \mathbb{Z}_{p}$ and sets $h=g^ag^\gamma$ and $h_1=g^{a^\eta}=g^{a\eta}$. For the corrupted authorities $A_k \in C_A$:

 \begin{itemize}
   \item The challenger generates $v_k, \beta_k, w_{k,j} ~\underleftarrow{~~R~~} \mathbb{Z}_{p}$ and sets $Y_k=e(g, g)^{v_k}$, $Z_k=g^{\beta_k}$ and $\{T_{k,j}=g^{w_{k,j}}\}_{a_{k,j} \in \widetilde{A_k}}$.
   \item Hence, the public-secret key pairs for $A_k \in C_A$ is given as $\left\{(v_{k}, \beta_k, [w_{k,1},\ldots,w_{k,n_k}]), (Y_{k}, Z_k, [T_{k,1},\ldots,T_{k,n_k}])\right\}$.
   \item Challenger provides the secret-public key pairs of the corrupted authorities to the adversary.
   \item Hence, the adversary can compute $D_{k,\omega}, D_{k,\omega}^1,$ and $D_{k,\omega}^j, \forall a_{k,j} \in \widetilde{A_\omega^k}$ himself for user $\omega$ without interacting with the challenger.
 \end{itemize}
 For the authorities who are not corrupted and not $\kappa$ (i.e., $A_k \notin C_A \cup \kappa$):

  \begin{itemize}
    \item Challenger generates $v_k, \beta_k, w_{k,j} ~\underleftarrow{~~R~~} \mathbb{Z}_{p}$ and sets $Y_k=e(g^b, g^{v_k})=e(g, g)^{bv_k}$, and $Z_k=g^{\beta_k}$.
    \item For the attributes ${a_{k,j} \in \widetilde{A_C} \cap \widetilde{A_k}}$, the challenger sets $T_{k,j}=g^{w_{k,j}}$.
    \item Other attributes i.e., ${a_{k,j} \in \widetilde{A_k} - \widetilde{A_C^k}}$, the challenger sets $T_{k,j}=h_1^{w_{k,j}}=g^{a \eta w_{k,j}}$.
    \item Hence, the public-secret key pairs for $A_k \notin C_A \cup \kappa$ is given as\\ $\left\{(bv_{k}, \beta_k, w_{k,1},\ldots,w_{k,n_k}), (Y_{k}, Z_k, T_{k,1},\ldots,T_{k,n_k})\right\}$.
    \item Challenger sends public keys $(Y_{k}, Z_k, T_{k,1},[T_{k,1},\ldots,T_{k,n_k}])$ to the adversary.
    \item Now authority $A_k$ assigns a polynomial $q_x$  with degree $d_x$ for every node in it's access tree $\mathbb{T}^k$. It first sets up a polynomial $q_x$ of degree $d_x$ for the root node $x$. It sets $q_x(0) = r_{k,\omega}$ and then sets rest of the points randomly to completely fix $q_x$. Now it sets polynomials for each child node $x'$  of $x$ as follows: $q_{x'}(0) = q_x(index(x'))$.
    \item Then computes $D_{k,\omega}=g^{-bv_{k}}g^{\frac{(a+\gamma)\beta_k}{r_{k,\omega}+\omega}}g^{\frac{a\eta r_{k,\omega}}{\beta_k+\omega}}, D_{k,\omega}^1=g^{\frac{a+\gamma}{r_{k,\omega}+\omega}},$ and $D_{k,\omega}^j=g^{\frac{a\eta q_{parent(a_{k,j})}(index(a_{k,j}))}{(\beta_k+\omega)w_{k,j}}}, \forall {a_{k,j} \in \widetilde{A_C} \cap \widetilde{A_k}}$ or $D_{k,\omega}^j=g^{\frac{ q_{parent(a_{k,j})}(index(a_{k,j})))}{(\beta_k+\omega)w_{k,j}}}, \forall {a_{k,j} \in \widetilde{A_k} - \widetilde{A_C^k}}$
  \end{itemize}
 If $A_k = \kappa$:
  \begin{itemize}
    \item Challenger generates $\beta_{\kappa}, w_{\kappa,j} ~\underleftarrow{~~R~~} \mathbb{Z}_{p}$.
     \item Challenger sets $Y_{\kappa}=e(g^a, g^b)\prod \limits_{j=1, j\neq \kappa}^n  Y_j^{-1}=e(g,g)^{{\left(ab-\sum \limits_{A_k \notin C_A \cup \kappa}bv_k-\sum \limits_{A_k \in CA}v_k\right)}}$, $Z_{\kappa}=g^{\beta_{\kappa}}$.
     \item For the attributes ${a_{\kappa,j} \in \widetilde{A_C} \cap \widetilde{A_\kappa}}$, the challenger sets $T_{\kappa,j}=g^{w_{\kappa,j}}$.
    \item Other attributes i.e., ${a_{\kappa,j} \in \widetilde{A_{\kappa}} - \widetilde{A_C^{\kappa}}}$, the challenger sets $T_{\kappa,j}=h_1^{w_{\kappa,j}}=g^{(a\eta)w_{\kappa,j}}$.
    \item Hence, the public-secret key pairs for $A_k = \kappa$ is given as $[(ab-\sum \limits_{A_k \notin C_A \cup \kappa}bv_k-\sum \limits_{A_k \in CA}v_k), \beta_k, w_{k,1},...,w_{k,n_k}, (Y_{\kappa}, Z_{\kappa}, T_{\kappa,1},...)]$.
    \item Challenger sends public keys $(Y_{\kappa}, Z_{\kappa}, T_{\kappa,1},...)$ to the adversary.

    \item Since $\mathbb{T}^{\kappa}(\widetilde{A_C^{\kappa}}\cap \widetilde{A_\omega^{\kappa}})=0$, it should be stressed that maximum $d_x-1$ number of  children of node $x$ could be satisfied. It first defines a polynomial $q_x$ of degree $d_x$ for the root node $x$ such that $q_x(0) = r'$.  For each satisfied child $x'$  of $x$, the procedure chooses a random point $e_{x'}$ and sets $q_x(index(x')) = e_{x'}$. We will show later in this section that it is possible to implicitly define remaining children nodes using polynomial interpolation.  Now it recursively defines polynomials for the rest of the nodes in the tree as follows. For each child node $x'$ of $x$, $q_{x'}(0)=q_x(index(x'))$.
    \item Computes $D_{\kappa,\omega}=g^{\left(-ab+\sum \limits_{A_k \notin C_A \cup k*}bv_k+\sum \limits_{A_k \in CA}v_k\right)}g^{\frac{(a+\gamma)\beta_{\kappa}}{(r+\omega)}}g^{\frac{a\eta r'}{\beta_{\kappa}+\omega}},\\ D_{\kappa,\omega}^1=g^{\frac{(a+\gamma)}{(r+\omega)}},$ and $D_{\kappa,\omega}^j=h_1^{\frac{q_{parent{a_{\kappa,j}}}(index(a_{\kappa,j}))}{w_{\kappa,j}}} \forall {a_{\kappa,j} \in A_C \cap \widetilde{A_k}}$ or
        $D_{\kappa,\omega}^j=g^{\frac{r'\Delta_{0, S}(a_{\kappa,j})+\sum \limits_{i=1}^{d_{\kappa}-1} e_i\Delta_{i, S}(a_{\kappa,j})}{w_{\kappa,j}}}, \forall {a_{\kappa,j} \in \widetilde{A_k} - \widetilde{A_C^k}}$ where $r'=r+b\frac{(\beta_{\kappa}+\omega)}{\eta}$.
  \end{itemize}
The following lemma proves $D_{\kappa,\omega}$ and $D_{\kappa,\omega}^j$ are correct and the challenger can generate these without knowing $a, b, c$ from $g^a, g^b, g^c$.\\

\noindent \textbf{Lemma 1.} \emph{$D_{\kappa,\omega}$ and $D_{\kappa,\omega}^j$ are correct and they can be generated by challenger without knowing $a, b, c$ from $g^a, g^b, g^c$.}

\noindent \textbf{Proof.}
In the beginning of the access structure we chosen $e_1$, $e_2$, $\ldots$, $e_{d_\kappa-1}$ random values for children nodes of root node and we set $q_x(0)=r'=r+b\frac{(\beta_{\kappa}+\omega)}{\eta}$. Hence, using the polynomial interpolation technique, we can implicitly assign values to other children nodes of root using the following valid polynomial function:
\begin{equation*}
    q_x=r'\Delta_{0, S}(x)+\sum \limits_{i=1}^{d_{\kappa}-1} e_i\Delta_{i, S}(x), S \in \{{0, 1, 2, \ldots, d_\kappa-1}\}.
\end{equation*}
During the key extraction, for the correctness, the challenger hides $r'$ in attribute $a_{\kappa,j} \in \{\widetilde{A_k} - \widetilde{A_C^k}\}$. Since, $T_{\kappa,j}=h_1^{w_{\kappa,j}}$, the $D_{\kappa,\omega}^j$ $\forall$ $a_{\kappa,j} \in \{\tilde{A}_k - \tilde{A}_C^k\}$ is given as
\begin{eqnarray*}
  D_{\kappa,\omega}^j &=& g^{\frac{q_{parent(a_{\kappa,j})}(index(a_{\kappa,j}))}{w_{\kappa,j}}}, \\
   &=& g^{\frac{r'\Delta_{0, S}(a_{\kappa,j})+\sum \limits_{i=1}^{d_{\kappa}-1} e_i\Delta_{i, S}(a_{\kappa,j})}{w_{\kappa,j}}},
\end{eqnarray*}
which is valid and identical to that in the original scheme. Now we will prove that the challenger can generate $D_{\kappa,\omega}^j$ and $D_{\kappa,\omega}$ for $a_{\kappa,j} \in \{\widetilde{A_k} - \widetilde{A_C^k}\}$ without knowing $a, b,$ and $c$.
\begin{eqnarray*}
  D_{\kappa,\omega}^j \!\!\!\! &=& g^{\frac{\left(r+b\frac{(\beta_{\kappa}+\omega)}{\eta}\right)\Delta_{0, S}(a_{\kappa,j})+\sum \limits_{i=1}^{d_{\kappa}-1} e_i\Delta_{i, S}(a_{\kappa,j})}{w_{\kappa,j}}},\\
  &=&g^{{\left(r+b\frac{(\beta_{\kappa}+\omega)}{\eta}\right)\frac{\Delta_{0, S}(a_{\kappa,j})}{w_{\kappa,j}}+\sum \limits_{i=1}^{d_{\kappa}-1} e_i\frac{\Delta_{i, S}(a_{\kappa,j})}{w_{\kappa,j}}}},\\
  &=&\left[g^r{\left(g^b\right)}^{\left(\frac{\beta_{\kappa}+\omega}{\eta}\right)}\right]^{\frac{\Delta_{0, S}(a_{\kappa,j})}{w_{\kappa,j}}}\prod \limits_{i=1}^{d_{\kappa}-1} g^{\frac{e_i\Delta_{i, S}(a_{\kappa,j})}{w_{\kappa,j}}},
\end{eqnarray*}
and
\begin{eqnarray*}
      \nonumber D_{\kappa,\omega} \!\!\!\!\!\!  &=&\!\!\!\!\!\!  g^{\left(-ab+\sum \limits_{A_k \notin C_A \cup \kappa}bv_k+\sum \limits_{A_k \in CA}v_k\right)}g^{\frac{(a+\gamma)\beta_{\kappa}}{(r+\omega)}}g^{\frac{a\eta r'}{\beta_{\kappa}+\omega}} \\
   \nonumber &=&\!\!\!\!\!\!  g^{\left(-ab+\sum \limits_{A_k \notin C_A \cup \kappa}bv_k+\sum \limits_{A_k \in CA}v_k\right)}g^{\frac{(a+\gamma)\beta_{\kappa}}{(r+\omega)}}g^{\frac{a\eta (r+b\frac{(\beta_{\kappa}+\omega)}{\eta})}{\beta_{\kappa}+\omega}}\\
     &=&\!\!\!\!\!\!  g^{-ab}g^{\left(\sum \limits_{A_k \notin C_A \cup \kappa}bv_k+\sum \limits_{A_k \in CA}v_k\right)}g^{\frac{(a+\gamma)\beta_{\kappa}}{(r+\omega)}}g^{\frac{a\eta r}{\beta_{\kappa}+\omega}}g^{ab}\\
      &=& \!\!\!\!\!\! g^{\left(\sum \limits_{A_k \notin C_A \cup \kappa}bv_k+\sum \limits_{A_k \in CA}v_k\right)}g^{\frac{(a+\gamma)\beta_{\kappa}}{(r+\omega)}}g^{\frac{a\eta r}{\beta_{\kappa}+\omega}}\\
 &=&\!\!\!\!\!\!{\left(g^b\right)}^{\left(\sum \limits_{A_k \notin C_A \cup \kappa}v_k\right)}g^{\left(\sum \limits_{A_k \in CA}v_k\right)}h^{\frac{\beta_{\kappa}}{r+\omega}}h_1^{\frac{ r}{\beta_{\kappa}+\omega}}~~~~~ \blacksquare
                          \end{eqnarray*}
Once adversary received all the credential from the challenger, he will send two messages, $m_0$ and $m_1$ to the challenger. Now the challenger randomly chooses one of the messages  and encrypts it and sends the encrypted message back to the adversary. Let us denote the message chosen by the challenger as $m_b$ where $b=0$ or $1$ and the encrypted message as $C = \{ C_1 = m_b.Z, C_2= g^{c}, C_3=\prod\limits_{k \in A_C}{\left(g^{c}\right)}^{\beta_k} \textrm{and}
\left \{ C_{k,j} = {\left(g^{c}\right)}^{w_{k,j}} \right \}_{\forall k\in I_{C}, a_{k,j} \in \tilde{A}_m^j} \}$. We stress here that $C$ is a valid encryption of the message $m_b$ if $Z=e(g,g)^{abc}$. Hence, the adversary should have his usual non-negligible advantage $\upsilon$ of correctly identifying the message $m_b$. However, when $Z\neq e(g,g)^{abc}$, then $C$ is just random value, hence, the adversary can have no more than $\frac{1}{2}$ probability of guessing correctly. Hence, if the adversary guesses correctly then challenger guesses that $Z=e(g,g)^{abc}$  and if adversary is wrong then challenger guesses that $Z\neq e(g,g)^{abc}$, hence, the challenger has an advantage of $\frac{\upsilon}{2}$ in distinguishing whether $Z= e(g,g)^{abc}$. Hence, an adversary who breaks our scheme with advantage $\upsilon$ implies an algorithm for breaking DBDH assumption with non-negligible advantage $\frac{\upsilon}{2}$. We can conclude that the proposed scheme is selective ID secure.\\

\begin{figure*}
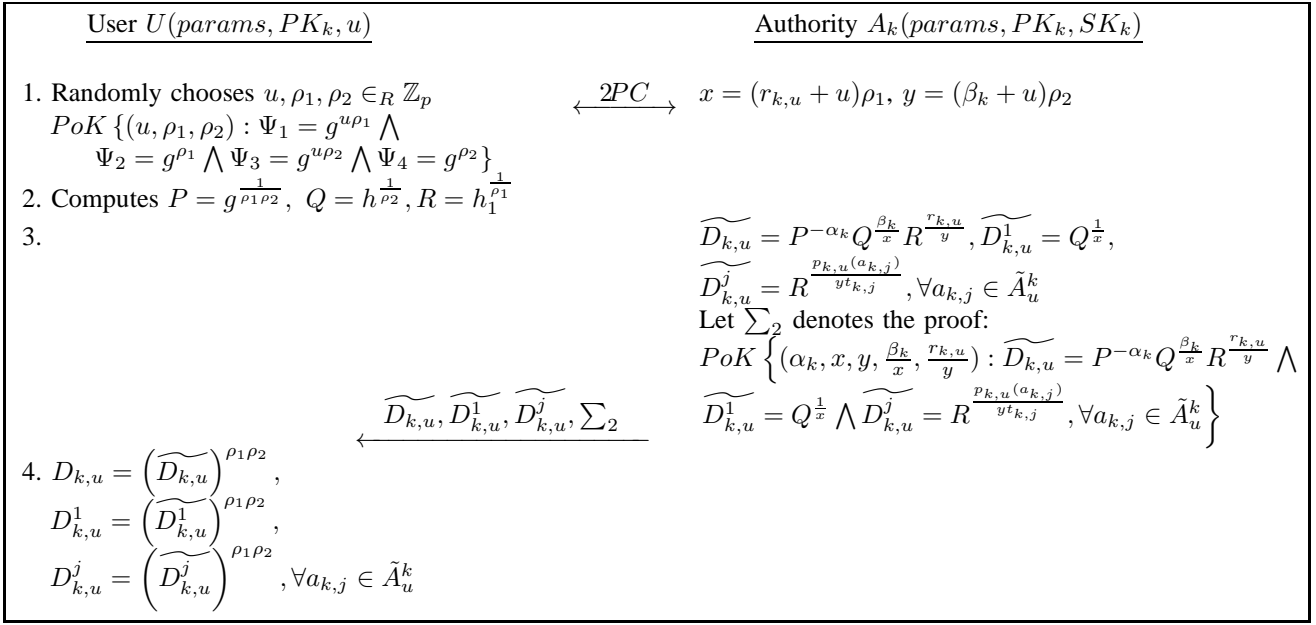

\setlength{\unitlength}{0.14in} 
\framebox{
\parbox[t][8.0cm]{17cm}{
\centering 
\makebox[9cm][l]{\ ~~~~~~\underline{User $U(params, PK_k,u)$}}\makebox[8cm][l]{\ ~~~~~\underline{Authority $A_k(params, PK_k, SK_k)$}}\vspace{0.5cm}\\
\makebox[9cm][l]{1. Randomly chooses $u, \rho_1, \rho_2 \in_R \mathbb{Z}_p$ $~~~~~~~~~~~~~~\underleftarrow{~~~2}\!\underrightarrow{PC~~~}$}\makebox[10cm][l]{$x = (r_{k,u} + u)\rho_1$, $y = (\beta_{k} + u)\rho_2$}\\
\makebox[9cm][l]{   ~~$PoK\left\{(u, \rho_1,\rho_2):\Psi_1=g^{u\rho_1}\bigwedge  \right.$}\makebox[8cm][l]{}\\
\makebox[9cm][l]{$~~~~~~~\left.\Psi_2=g^{\rho_1}\bigwedge\Psi_3=g^{u\rho_2}\bigwedge\Psi_4=g^{\rho_2}\right\}$}\makebox[8cm][l]{}\\
\makebox[9cm][l]{2. Computes $P = g^{\frac{1}{\rho_1\rho_2}},~Q = h^{\frac{1}{\rho_2}}, R = h_1^{\frac{1}{\rho_1}}$}\makebox[8cm][l]{}\\
\makebox[9cm][l]{3. }\makebox[8cm][l]{$\widetilde{D_{k,u}}=P^{-\alpha_k}Q^{\frac{\beta_k}{x}}R^{\frac{r_{k,u}}{y}}, \widetilde{D_{k,u}^{1}} = Q^{\frac{1}{x}},$}\\
\makebox[9cm][l]{ }\makebox[8cm][l]{$\widetilde{D_{k,u}^j} = R^{\frac{p_{k,u}(a_{k,j})}{yt_{k,j}}}, \forall a_{k,j} \in \tilde{A}_u^k$}\\
\makebox[9cm][l]{ }\makebox[8cm][l]{Let $\sum_2$ denotes the proof:}\\
\makebox[9cm][l]{ }\makebox[8cm][l]{$PoK \left\{(\alpha_k, x, y, \frac{\beta_k}{x}, \frac{r_{k,u}}{y}):\widetilde{D_{k,u}}=P^{-\alpha_k}Q^{\frac{\beta_k}{x}}R^{\frac{r_{k,u}}{y}}\bigwedge \right. $} \\
\makebox[9cm][l]{$~~~~~~~~~~~~~~~~~~~~~~~~~~~~~~~~~~~~\underleftarrow{~~~\widetilde{D_{k,u}}, \widetilde{D_{k,u}^{1}}, \widetilde{D_{k,u}^j}, \sum_2~~~}$ }\makebox[8cm][l]{$\left.\widetilde{D_{k,u}^{1}} = Q^{\frac{1}{x}} \bigwedge  \widetilde{D_{k,u}^j} = R^{\frac{p_{k,u}(a_{k,j})}{yt_{k,j}}}, \forall a_{k,j} \in \tilde{A}_u^k\right\}$}\\
\makebox[9cm][l]{4. $D_{k,u}=\left(\widetilde{D_{k,u}}\right)^{\rho_1\rho_2},$}\makebox[8cm][l]{}\\
\makebox[9cm][l]{ $~~D_{k,u}^{1}=\left(\widetilde{D_{k,u}^{1}}\right)^{\rho_1\rho_2},$}\makebox[8cm][l]{}\\
\makebox[9cm][l]{ $~~D_{k,u}^j = \left(\widetilde{D_{k,u}^j}\right)^{\rho_1\rho_2}, \forall a_{k,j} \in \tilde{A}_u^k$}\makebox[8cm][l]{}\\
}}
\caption{Anonymous key issuing protocol for decentralized KP-ABE}
\label{Anonymous key issuing protocol}
\end{figure*}

\noindent \textbf{Theorem 2.} \emph{Let $\xi$ be a $(t, q_S, \epsilon)$-selective identity secure IBE system (IND-sID-CPA). Suppose $\xi$
admits $N$ distinct identities. Then $\xi$ is also a $(t, qS, N\epsilon)$-fully secure IBE (IND-ID-CPA).}\\

\noindent \textbf{Proof.} It has been shown by Boneh et. al in [34; see Theorem 7.1.] that, if any selective identity model is secure under DBDH assumption then it is also fully secure (i.e., adaptive identity). Consequently, if the system has sufficiently high selective-identity security (which requires using a bilinear group of sufficiently large size p) then the system is also a fully secure with adequate security [34]. This means that the selective-ID secure system proposed in this paper is fully secure system in its own right, assuming we use a large enough group so that the DBDH problem is sufficiently difficult.

\section{Anonymous Key Issuing Protocol}
As described by Chase et. al., the ABE system comprises of several types of authorities where some authorities require public identities such as name or social security number of the users to issue credentials \cite{imaabe}. In these cases the users will need to identify themselves in any case in order to obtain the decryption keys for a specific set of attributes. However, there are many attributes, which do not belong to this category e.g., ability to drive a car. However, one should be able to prove the ability to do something in an examination and then get the corresponding credential, without presenting any identifying information.

In an anonymous credential system [30,31,32], the users can obtain and prove the possession of credentials while remaining anonymous. In such work it is assumed that each user has a unique secret key. We can assume that this secret key could be the user’s GID. Then the user can interact with each authority under a different pseudonym (generated using GID) in such a manner that it is impossible to link multiple pseudonyms belonging to the same user. There are two efficient secure protocols in literature to prove that a given secret key is valid and to prevent users from loaning out their keys [30,31,32]: public-key infrastructure (PKI)-assured non-transferability and all-or-nothing non-transferability. In PKI-assured non-transferability, sharing a key implies also sharing a particular, valuable secret key from outside the system (e.g., the secret key that gives access to the user's bank account). In all-or-nothing non-transferability, sharing just one pseudonym or credential implies sharing all of the user's other credentials and pseudonyms in the system, i.e., sharing all of the user's secret keys inside the system.

In the proposed work, we exploited anonymous credentials to allow the users to obtain decryption keys from the authorities without revealing their GID’s. The basic idea is to let the GID play the role of the anonymous credential secret key. The user obtains anonymous credentials as explained earlier. When the user wants to obtain decryption keys corresponding to a set of attributes, he proves (via the anonymous credential system) that he is the owner of a credential for these attributes. Then he uses the ABE system to obtain decryption keys.

In particular, in the proposed key issuing protocol, we coupled the properties of anonymous credential system with ABE. Hence, the user can obtain a set of decryption keys for his secret GID without revealing any information about that GID to the authority (preserve the privacy). At the same time, the authority is guaranteed that the agreed upon decryption-keys are the only information that the user learns from the transaction (i.e., provide security by mitigating user collusion).

The anonymous key issuing protocol preserves the user's identities from the authorities. Hence, the authorities do not know the user's GID nor can cause failures for certain GID's. This concept is from blind IBE schemes  \cite{IBE1}. We define this algorithm as $\left[U(params, PK_k,u, GID, decom)\leftrightarrows\right.$ $\left. A_k(params, PK_k, SK_k)\right]\rightarrow Decryption Keys for User$.
In this algorithms, the user runs \textit{Commit} and sends the \textit{com} to authority and keeps the \textit{decom}. Then the user and authority interact with each other to generate decryption keys. The decryption keys for the user will be  generated only if the output of \textit{Decommit} is $1$.

From \cite{IBE1}, the anonymous key extraction protocol should satisfy the following two properties: leak-freeness and selective-failure blindness. Leak-freeness requires that by executing the anonymous key issuing algorithm with honest authorities, the malicious user cannot find out anything which it is supposed to know by executing the algorithm without privacy preservation with the authorities. Selective-failure blindness requires that malicious authorities cannot discover anything about the user�s identifier and his attributes, and cause the algorithm to selectively fail depending on the user�s identifier and his attributes. These two properties can be formalized by using the following games.

\subsection{Security Games}

\textit{Leak-Freeness.} This game is defined by a real experiment and an ideal experiment. In real experiment, the user (adversarial user) and authority interact with each other using the proposed anonymous key issuing protocol. Hence, the user will obtain randomized decryption keys. However the user will be able remove the randomness to get decryption keys. In an ideal experiment, there will be a simulator that will interact (on behalf of user) with trusted authority to obtain decryption keys (without randomization). The proposed algorithm is free from leak only if no efficient distinguisher can distinguish whether the user is executing real or ideal experiments.

\textit{Selective-failure Blindness.} In this game, the authority (adversarial) outputs pair of GIDs (i.e., $GID_0$ and $GID_1$). Then the adversary given with commitments $com_b$ and $com_{1-b}$ where $b \in \{0,1\}$ and black box access to $U(params, PK_k,GID_b, decom_b)$ and $U(params, PK_k,GID_{1-b}, decom_{1-b})$. The algorithm $U$ interact with the authority and output decryption keys i.e., $\left(SK_{u_b}, SK_{u_{1-b}}\right)$. The proposed algorithm is blind to selective-failure if no probabilistic polynomial-time adversary has non-negligible advantage over identifying $b$.

\subsection{Construction of anonymous key issuing protocol}

Fig.~ \ref{Anonymous key issuing protocol} shows the anonymous key issuing protocol where user with private value $u \in \mathbb{Z}_p$ and an authority $A_k$ with private values $r_{k,u}$, $\beta_k$ and $\alpha_k$ $\in \mathbb{Z}_p$ jointly computing decryption keys for user. The decryption keys for user $u$ are $D_{k,u}, D_{k,u}^{1}$ and $D_{k,u}^j$ $\forall a_{k,j} \in \widetilde{A_u^k}$. In order to obtain these keys, first user and $A_k$ interact with each other using two-party protocol (2PC). The 2PC protocol can be done via a general 2PC protocol for a simple arithmetic computation. Alternatively, we can do this more efficiently using the construction in \cite{2PC1}. The 2PC protocol takes $(u,\rho_1,\rho_2)~\in \mathbb{Z}_p$ from user and $(r_{k,u}, \beta_k)~\in \mathbb{Z}_p$ from $A_k$ and returns $x=(u+\beta_k)\rho_1 ~mod~p$ and $y=(u+r_{k,u})\rho_2 ~mod~p$ to $A_k$. Since, $\rho_1$ and $\rho_2$ were randomly generated by user, the authority $A_k$ cannot extract the user identity $u$ from $x$ and $y$. After executing the 2PC protocol, the user now computes $P = g^{\frac{1}{\rho_1\rho_2}},~Q = h^{\frac{1}{\rho_2}}$ and $R = h_1^{\frac{1}{\rho_1}}$ and send those values to $A_k$. Now $A_k$ computes $\widetilde{D_{k,u}}$, $\widetilde{D_{k,u}^{1}}$ and $\widetilde{D_{k,u}^j}~\forall a_{k,j}~\in~\widetilde{A_u^k}$ (i.e., randomized decryption credentials) using $P,~Q,~R,~x$ and $y$ and send them back to the user. Now the user exponentiates the obtained values by $\rho_1\rho_2$ to get the decryption keys. Note that, since $u$  coupled non-linearly within the decryption keys,  user collusion is not possible \cite{imaabe}.

\subsection{Security Analysis}
To obtain the decryption credential blindly from the authority $A_k$ $\forall k$, the user needs to prove that he holds the identifier $u$ in zero knowledge. As shown in Fig.~ \ref{Anonymous key issuing protocol}, the user randomly generates $\rho_1, \rho_2 \in_R \mathbb{Z}_p$ and computes $\Psi_1=g^{u\rho_1}, \Psi_2=g^{\rho_1}, \Psi_3=g^{u\rho_2},$ and $\Psi_4=g^{\rho_2}$ as commitments. At the end of the 2PC protocol, the authority obtains $x = (r_{k,u} + u)\rho_1~mod~p$, $y = (\beta_{k} + u)\rho_2~mod~p$. Then authority verifies $g^x\stackrel{?}{=}AB^{r_{k,u}}$ and $g^y\stackrel{?}{=}CD^{r_{k,u}}$. If they are correctly verified then the authority continues otherwise it aborts.

Now authority needs to proof that he knows $(\alpha_k, \frac{\beta_k}{x}, \frac{r_{k,u}}{y})$ in zero knowledge to the user. This will be done using the following steps:
\begin{enumerate}
  \item[1.] $A_k$ randomly generates $b_1, b_2, b_3 \in_R \mathbb{Z}_p$, computes $\widetilde{\widetilde{D_{k,u}}}=P^{-b_1}Q^{b_2}R^{b_3}$ and sends $\widetilde{\widetilde{D_{k,u}}}$, and ${\widetilde{D_{k,u}}}$ to the user
  \item[2.] User generates $c_1 \in_R \mathbb{Z}_p$ and sends it to the authority
  \item[3.] Authority computes $b_1'=b_1-c_1\alpha_k, b_2'=b_2-c_1\frac{\beta_k}{x}, b_3'=b_3-c_1\frac{r_{k,u}}{y}$ and sends $b_1', b_2', b_3'$ to the user
  \item[4.] User verifies $\widetilde{\widetilde{D_{k,u}}}\stackrel{?}{=}P^{-b_1'}Q^{b_2'}R^{b_3'}{\widetilde{D_{k,u}}}^{c_1}$, otherwise aborts
\end{enumerate}
We ignored the zero-knowledge proofs for  $\frac{1}{x}$ and $\frac{1}{y}$ for brevity since they are similar to the above proof.\\

\noindent \textbf{Theorem 3.} \emph{The proposed anonymous key issuing protocol is both leak-free and selective-failure blind.}\\

\noindent \textbf{Proof.} \textbf{Leak freeness.}  Suppose there exists an adversary $\mathfrak{U}$ in the real experiment (where $\mathfrak{U}$ is interacting with an honest authority $A_k$ running the anonymous key issuing protocol) and a simulator $\mathfrak{\tilde{U}}$ in the
ideal experiment (where $\mathfrak{\tilde{U}}$ can access the trusted authority running the key issuing protocol without privacy preservation) such that no efficient distinguisher $\mathfrak{D}$ can distinguish the real experiment from the ideal experiment. The simulator $\mathfrak{\tilde{U}}$ simulates the communication between the distinguisher $\mathfrak{D}$ and the adversary $\mathfrak{U}$ by passing the input of $\mathfrak{D}$ to $\mathfrak{U}$ and the output of $\mathfrak{U}$ to $\mathfrak{D}$.  The simulator $\mathfrak{\tilde{U}}$ works as follows:
\begin{enumerate}
  \item[1.] $\mathfrak{\tilde{U}}$ sends the adversary $\mathfrak{U}$ the public-key $PK_k$ of $A_k$
  \item[2.] The adversary $\mathfrak{U}$ must proof the possession of $u$ in zero-knowledge to $\mathfrak{\tilde{U}}$. If proof is successful then $\mathfrak{\tilde{U}}$ obtains $(u, \rho_1, \rho_2)$ using rewind technique
  \item[3.] $\mathfrak{\tilde{U}}$ sends $u$ to the trusted party. The trusted party runs KeyGen to generates ($D_{k,u}$, $D_{k,u}^1$, $D_{k,u}^j$)  and responds to $\mathfrak{\tilde{U}}$
  \item[4.] Now $\mathfrak{\tilde{U}}$  computes $\left({D_{k,u}}^{\frac{1}{\rho_1\rho_2}}, {D_{k,u}^1}^{\frac{1}{\rho_1\rho_2}}, {D_{k,u}^j}^{\frac{1}{\rho_1\rho_2}}\right)$ and sends them to $\mathfrak{U}$
\end{enumerate}
If ($D_{k,u}$, $D_{k,u}^1$, $D_{k,u}^j$) are correct keys from the trusted authority in the ideal experiment, then $\left({D_{k,u}}^{\frac{1}{\rho_1\rho_2}}, {D_{k,u}^1}^{\frac{1}{\rho_1\rho_2}}, {D_{k,u}^j}^{\frac{1}{\rho_1\rho_2}}\right)$ are the correct keys from $A_k$ in the real experiment. Hence, ($D_{k,u}$, $D_{k,u}^1$, $D_{k,u}^j$) and $\left({D_{k,u}}^{\frac{1}{\rho_1\rho_2}}, {D_{k,u}^1}^{\frac{1}{\rho_1\rho_2}}, {D_{k,u}^j}^{\frac{1}{\rho_1\rho_2}}\right)$ are correctly distributed and no efficient distinguisher can distinguish the real experiment with the ideal experiment.\\

\noindent \textbf{Proof.} \textbf{Selective-failure blindness.} The adversarial authority $A_k$ submits the public key $PK_k$, and two GIDs $u_0$ and $u_1$. Then, a bit $b \in \{0, 1\}$ is randomly selected. $A_k$ can have a black box access to $u_0$'s and $u_1$'s parameters i.e., $U(params, PK_k, u_b)$ and $U(params, PK_k, u_{b-1})$. Then, $U$ executes the anonymous key issuing protocol with $A_k$ and outputs secret keys for $u_b$ and $u_{1-b}$ i.e., $SK_{u_b}$ and $SK_{u_{1-b}}$.
\begin{itemize}
  \item[1.] If $SK_{u_b}\neq \perp$ and $SK_{u_{1-b}} \neq \perp$ then $A_k$ is given ($SK_{u_b}$, $SK_{u_{1-b}}$)
  \item[2.] If $SK_{u_b}\neq \perp$ and $SK_{u_{1-b}} = \perp$ then $A_k$ is given ($\epsilon$, $\perp$)
  \item[3.] If $SK_{u_b}= \perp$ and $SK_{u_{1-b}} \neq \perp$ then $A_k$ is given ($\perp$, $\epsilon$)
  \item[4.] If $SK_{u_b} = \perp$ and $SK_{u_{1-b}} = \perp$ then $A_k$ is given ($\perp$, $\perp$)
\end{itemize}
At the end $A_k$ submits his prediction on $b$.

In the anonymous key issuing protocol, $U$ sends four random parameters $\Psi_1=g^{u_b\rho_1} \in \mathbb{G}_{1}, \Psi_2=g^{\rho_1} \in \mathbb{G}_{1}, \Psi_3=g^{u_b\rho_2} \in \mathbb{G}_{1},$ and $\Psi_4=g^{\rho_2} \in \mathbb{G}_{1}$ to the adversarial authority $A_k$ and proves $PoK{(u_b, \rho_1,\rho_2)}$. Now it is $A_k$'s turn to respond. So far, $A_k$�s view on the two black boxes is computationally undistinguishable. Otherwise, the hiding property of the commitment scheme and the witness undistinguishable property of the zero-knowledge proof will be broken. Suppose that $A_k$ uses any computing strategy to output secret keys \{$D_{k,u_b}, D_{k,u_b}^1, D_{k,u_b}^j \forall a_{k,j} \in \tilde{A}_{u_b}^k$\} for the first black box. In the following, we will show that $A_k$ can predict $SK_{u_b}$ of $U$ without interacting with the two black boxes:

\begin{itemize}
  \item[1.]  $A_k$ checks
  \item[] $PoK \left\{(\alpha_k, \frac{\beta_k}{x}, \frac{r_{k,u}}{y}):\widetilde{D_{k,u}}=P^{-\alpha_k}Q^{\frac{\beta_k}{x}}R^{\frac{r_{k,u}}{y}}\bigwedge \right. $
  \item[] $\left.\widetilde{D_{k,u}^{1}} = Q^{\frac{1}{x}} \bigwedge  \widetilde{D_{k,u}^j} = R^{\frac{p_{k,u}(a_{k,j})}{yt_{k,j}}}, \forall a_{k,j} \in \tilde{A}_u^k\right\}$
  \item[] If proof fails, $A_k$ sets $SK_{0} = \perp$
  \item[2.] $A_k$ generates different secret keys \{$D_{k,u_b}, D_{k,u_b}^1, D_{k,u_b}^j \forall a_{k,j} \in \tilde{A}_{u_b}^k$\} for the second black box and a proof of knowledge:
      \item[] $PoK \left\{(\alpha_k, \frac{\beta_k}{x}, \frac{r_{k,u}}{y}):\widetilde{D_{k,u}}=P^{-\alpha_k}Q^{\frac{\beta_k}{x}}R^{\frac{r_{k,u}}{y}}\bigwedge \right. $
  \item[] $\left.\widetilde{D_{k,u}^{1}} = Q^{\frac{1}{x}} \bigwedge  \widetilde{D_{k,u}^j} = R^{\frac{p_{k,u}(a_{k,j})}{yt_{k,j}}}, \forall a_{k,j} \in \tilde{A}_u^k\right\}$
  \item[] If proof fails, $A_k$ sets $SK_{1} = \perp$
  \item[3.] Finally $A_k$ outputs his prediction on ($u_0$, $u_1$) with ($SK_{u_b}$, $SK_{u_{1-b}}$) if $SK_{u_b}\neq \perp$ and $SK_{u_{1-b}} \neq \perp$; ($\epsilon$, $\perp$) if $SK_{u_b}\neq \perp$ and $SK_{u_{1-b}} = \perp$; ($\perp$, $\epsilon$) if $SK_{u_b}= \perp$ and $SK_{u_{1-b}} \neq \perp$; ($\perp$, $\perp$) if $SK_{u_b} = \perp$ and $SK_{u_{1-b}} = \perp$.
\end{itemize}

\begin{table*}[!ht]\caption{Comparison of Computational Cost and Ciphertext length}
  \centering
\begin{tabular}{|c|c|c|c|c||c|}
  \hline
                                &\!\!\!\! Authority Set.\!\!\!\!   & Key Generation& Encryption    & Decryption            & $|$Ciphertext$|$ \\ \hline
   \multirow{2}{*}{\!\!\!\!\!\!\!\!\!\!Our Scheme\!\!\!\!\!\!\!\!\!\!}    & \multirow{2}{*}{$(nN+2N)C_e$}      & $(5N+2nN)C_m+$& $(2N-1)C_m+$  &\!\!\!\!\!\!\!\!\!\!$(1+N+nN)C_m+nNC_e$\!\!\!\!\!\!   &$(2+nN)|G_1|$  \\
                                &                   & $(4N+nN)C_e$  & $(1+2N+nN)C_e$&$+(1+N+nN)C_p\!\!\!\!$          & $+|G_2|$ \\ \hline
  {Han et. al. }      & \multirow{2}{*}{$(nN+2N)C_e$}      & $(2N+nN)C_m+$ &$(2N-1)C_m+$   & $(N+2)C_m+(1+nN)C_e$ & $(2+nN)|G_1|$ \\
   \cite{main1}                 &                   &$(3N+nN)C_e$   &$(1+2N+nN)C_e$ & $+(1+N+nN)C_p$         &$+|G_2|$  \\ \hline
 {Chase  }           & \multirow{2}{*}{$(nN+1)C_e$}       & \multirow{2}{*}{$nNC_m+nNC_e$} &\multirow{2}{*}{$C_m+(2+nN)C_e$}& $(nN+2)C_m+$          & $(1+nN)|G_1|$ \\
       \cite{maabe}                         &                   &               &               & $nNC_e+(nN+1)C_p$     & $+|G_2|$ \\ \hline
  {\!\!\!\!Chase et. al. \!\!\!\!\!\!}   & $N(N-1)C_m+$ &$3N(N-1)C_m+$ & $C_m+(2+nN)C_e$ & $2C_m+nNC_e+$     &$(1+nN)|G_1|$  \\
     \cite{imaabe}                & $(nN+2N)C_e$      &$(nN+n^2N^2)C_e$ &                 & $(1+nN)C_p$       &  $+|G_2|$\\
  \hline
\end{tabular}
\label{Table: Comparision}
\end{table*}

The predication on ($u_0$, $u_1$) is correct, and has the identical distribution with the black box. Because $A_k$ performs the same check as the honest $U$, it outputs the valid keys as $U$ obtains from anonymous key issuing protocol. Hence, if $A_k$ can predict the final outputs of
the two black boxes, the advantage of $A_k$ in distinguishing the two black boxes is the same without the final outputs. Therefore, the advantage
of $A_k$ should come from the received $\Psi_1, \Psi_2, \Psi_3, \Psi_4 \in \mathbb{G}_{1}$ and $PoK{(u, \rho_1,\rho_2)}$. From the hiding property of the commitment scheme and witness undistinguishable property of the zero-knowledge proof, $A_k$ cannot distinguish one from the other with nonnegligible advantage. Hence our anonymous key issuing protocol is secure against selective-failure.

\section{Complexity Comparison}
In this section, we compare the computational complexity of our scheme against the following KP-ABE  schemes: Han et. al decentralized KP-ABE scheme \cite{main1}, Chase MA-ABE KP-ABE scheme with central authority \cite{maabe}, and Chase et.al MA-ABE KP-ABE scheme without central authority \cite{imaabe}. Let us assume there are $N$ number of attribute authorities and each authority monitors $n$ number of attributes. In order to compare the complexity at worst case scenario, let us assume that the ciphertext encrypted using all the attributes in the system (i.e., $nN$) and user has decryption credentials for all the attributes.  Let us denote the computational time (in ms) for one multiplication, one exponentiation,  and  one pairing as  $C_{m}$, $C_{e}$, and $C_{p}$, respectively.

 The Table \ref{Table: Comparision} compares the complexities in all four sub-algorithms and length of ciphertext. From the Table \ref{Table: Comparision}, it is obvious that the centralized algorithms in  \cite{maabe,imaabe} have lower computational complexity than the decentralized schemes proposed in this paper and in  \cite{main1}. However, since the algorithms in \cite{maabe,imaabe} need a central authority and/or interaction among authorities, they are vulnerable for single point of failure. Since the number of pairing operations (i.e., $C_p$) equal in both the proposed and  \cite{main1}, they share same order of computational complexity. However, the proposed algorithm mitigates the user collusion vulnerability  compared to \cite{main1}. Finally, the size of the ciphertext in all four algorithms are almost equal.

\section{Conclusions}
In this paper, we proposed a privacy-preserving key-policy attribute-based encryption scheme for a distributed data sharing environment. The proposed scheme enables users to download and decrypt the data from online such as cloud without revealing their attributes to the third-parties. The novelty of the work is to mitigate the user collusion attack in the existing scheme. We used anonymous key issuing protocol to strengthen the bind between user identity and decryption keys, hence,  two or more users cannot pool their keys to generate decryption keys for an unauthorized user. We validated the security of the proposed scheme using the decisional bilinear Diffie-Hellman standard complexity assumption.
\balance
%

\end{document}